\documentclass[aps,superscriptaddress,prb,amsmath,amssymb,reprint,longbibliography]{revtex4-2}

\usepackage[USenglish]{babel}
\usepackage[T1]{fontenc}
\usepackage[utf8]{inputenc}
\usepackage{graphicx}
\usepackage{upgreek}
\usepackage[unicode]{hyperref}
\usepackage{multirow}
\usepackage{verbatim} 
\usepackage[usenames, dvipsnames]{color} 

\begin{document}
	
\title{Hole spin coherence in InAs/InAlGaAs self-assembled quantum dots\\ emitting at telecom wavelengths}

\date{\today}

\author{E.~Evers}
\affiliation{Experimentelle Physik 2, Technische Universit\"at Dortmund, 44221 Dortmund, Germany}
	
\author{N.~E.~Kopteva}
\affiliation{Experimentelle Physik 2, Technische Universit\"at Dortmund, 44221 Dortmund, Germany}

\author{V.~Nedelea}
\affiliation{Experimentelle Physik 2, Technische Universit\"at Dortmund, 44221 Dortmund, Germany}

\author{A.~Kors}
\affiliation{Institute of Nanostructure Technologies and Analytics (INA), CINSaT, University of Kassel, D-34132 Kassel, Germany}

\author{R.~Kaur}
\affiliation{Institute of Nanostructure Technologies and Analytics (INA), CINSaT, University of Kassel, D-34132 Kassel, Germany}

\author{J.~P.~Reithmaier}
\affiliation{Institute of Nanostructure Technologies and Analytics (INA), CINSaT, University of Kassel, D-34132 Kassel, Germany}

\author{M.~Benyoucef}
\email{email: m.benyoucef@physik.uni-kassel.de}
\affiliation{Institute of Nanostructure Technologies and Analytics (INA), CINSaT, University of Kassel, D-34132 Kassel, Germany}

\author{M.~Bayer}
\affiliation{Experimentelle Physik 2, Technische Universit\"at Dortmund, 44221 Dortmund, Germany}

\author{A.~Greilich}
\email{email: alex.greilich@tu-dortmund.de}
\affiliation{Experimentelle Physik 2, Technische Universit\"at Dortmund, 44221 Dortmund, Germany}

\begin{abstract}
We report measurements of the longitudinal and transverse spin relaxation times of holes in an ensemble of self-assembled InAs/InAlGaAs quantum dots (QDs), emitting in the telecom spectral range. The spin coherence of a single carrier is determined using spin mode-locking in the inhomogeneous ensemble of QDs. Modeling the signal allows us to extract the hole spin coherence time to be in the range of $T_2 = 0.02 - 0.4\,\mu$s. The longitudinal spin relaxation time $T_1=0.5\,\mu$s is measured using the spin inertia method.
\end{abstract}

\maketitle

\section{Introduction}
\begin{figure*}[t]
	\begin{center}
		\includegraphics[trim=0mm 0mm 0mm 0mm, clip, width=2.15\columnwidth]{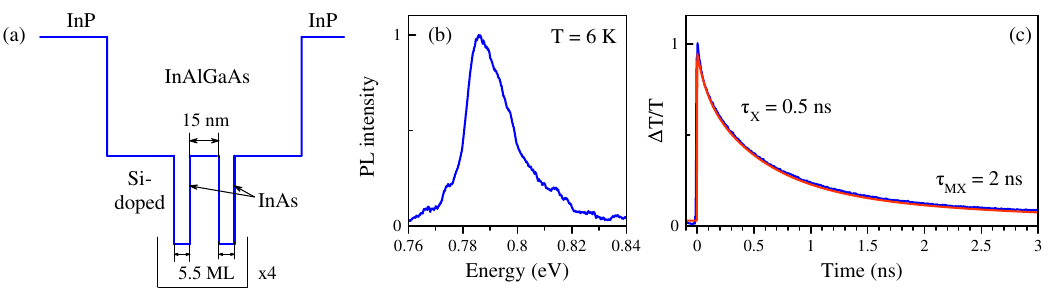}
		\caption{(a) Schematic of the potential profile and the sample structure. (b) Photoluminescence spectrum measured at $T = 6$\,K. (c) Dynamics of differential transmission ($\Delta T/T$) excited and detected at the energy of $E_\text{p} = 0.784$\,eV (blue). The red line is fit by a biexponential decay with $\tau_\text{X} = 0.5$\,ns and $\tau_\text{MX} = 2$\,ns. Pump and probe powers are 17\,mW and 1\,mW, respectively.}
		\label{Opt_prop}
	\end{center}
\end{figure*}

Quantum repeaters (QRs) are critical for achieving end-to-end security in fiber optics-based quantum communication networks~\cite{BriegelPRL98}. They allow the entanglement of fiber optic sections through quantum mechanical coupling or quantum teleportation, eliminating the need for additional security measures at the nodes~\cite{ChouScience07,YuanNature08}.

Implementing QRs requires quantum storage devices with sufficiently long coherence times to enable the stored state's entanglement with a photon for subsequent transmission~\cite{BhaskarNature20,BradleynpjQI22}. For this purpose, semiconductor quantum dots (QDs) seem to be an attractive candidate. Entanglement of a carrier spin with an emitted photon was demonstrated in QDs~\cite{GaoNature12,DeGreveNature12}, as well as an entanglement operation of two remote spins induced by a measurement of two indistinguishable photons~\cite{StockillPRL17}. The listed investigations focused mostly on QDs with an optical excitation wavelength of approximately 0.9\,$\mu$m. However, to be of practical relevance for fiber optics networks, QDs must operate in the C-telecom band, around 1.55\,$\mu$m.

Recent years have witnessed substantial advancements in single photon emission~\cite{Benyoucef13,PaulAPL17,OlbrichAPL17,Musial20,Musial21,LioAQT22,Katsumi_2023} and coherent manipulation of spin states for QDs operating in the telecom wavelength range~\cite{Rudno21,Podemski21,DusanowskiNatComm22}. While it has been established that spins can be efficiently oriented and manipulated through optical pulses, a critical aspect, the spin coherence time, representing the relevant information lifetime, has remained unexplored.

In this paper, we close this knowledge gap and provide measurements of the longitudinal and transverse spin relaxation times of holes in an ensemble of InAs self-assembled QDs emitting at telecom wavelengths.

\section{Experimental details}

The investigated QD sample is grown by molecular-beam epitaxy on a (100)-oriented InP substrate. It consists of 5.5\,nm InAs monolayers separated by InAlGaAs barriers. The sample schematic is presented in Fig.~\ref{Opt_prop}(a). The bottom barrier contains a Si $\delta$-doped layer at a distance of 15\,nm from the QD layer to provide resident electrons in the QDs. The density of QDs is approximately $10^{10}\,$cm$^{-2}$. We use eight QD layers separated by a barrier of 15\,nm to additionally study the implications of QD molecular states created by carriers tunneling between the neighboring QDs. Molecule states can be better protected from parameter fluctuations like charge or magnetic field variations~\cite{CarterPRL22,BoppAQT22}.

The sample is placed in helium gas for measurements at $T = 6-60$\,K. Figure~\ref{Opt_prop}(b) shows the photoluminescence (PL) spectrum of the QD ensemble measured at $T = 6$\,K. The PL spectrum has its maximum at around 0.784\,eV and shows inhomogeneously broadened emission with a width of about 20\,meV due to the spread of QD parameters.

Magnetic fields up to 4\,T are applied either transverse to the $\textbf{k}$-wavevector of light (Voigt geometry, $B_\text{V}$) or longitudinally to this direction (Faraday geometry, $B_\text{F}$). 

The spin coherence of spin-polarized carriers is measured by pump-probe Faraday rotation~\cite{Yugova2009}. To excite the QDs with telecom wavelengths, we use a pulsed Ti:Sapphire laser pumping an optical parametric oscillator at a pulse repetition rate of 76.7\,MHz (pulse repetition period of $T_R = 13$\,ns). The spectral width of the laser is below 2\,nm (0.8\,meV) with a pulse duration of about 1.5\,ps. The laser beam is split into the pump and probe. The photon energies ($E_\text{p}$) of both pulses are the same. The circularly polarized pump pulses create the spin polarization, while the linearly polarized probe pulses detect the spin polarization via the Faraday rotation effect. To suppress the scattered light, we use a double-modulation technique. The pump beam is modulated between $\sigma^+$ and $\sigma^-$ by an electro-optical modulator at $f_\text{m} = 10$\,kHz. The probe beam is intensity modulated by a photo-elastic modulator at $100$\,kHz. The signal is amplified by a lock-in device operated at a differential frequency and recorded versus the time delay between pump and probe.

\section{Experimental results and discussion}

\subsection{Spin signal components}

Figure~\ref{Opt_prop}(c) shows the exciton population recombination dynamics measured by differential transmission ($\Delta T/T$) at the PL maximum $E_\text{p} = 0.784$\,eV. Linearly polarized pump pulses generate the exciton population, and its dynamics are monitored by the probe with orthogonal linear polarization. The $\Delta T/T$ is fitted by a double exponential decay function, giving the short decay time of $\tau_\text{X} =0.5$\,ns and the long decay time of $\tau_\text{MX} = 2$\,ns (X stands for the direct exciton, with electron and hole residing in the same QD, while MX - for an indirect, molecular exciton, with electron and hole in different QDs). The (maximum) barrier thickness of 15\,nm is considered thin enough for electrons to tunnel between the QDs~\cite{BrackerAPL07}. The short decay component is related to recombination of the direct excitons, while the longer one $\tau_\text{MX}=2$\,ns can be identified to indirect exciton recombination time. As was shown on the structures in Ref.~\cite{Belykh2015}, the uncoupled QD layers demonstrate only a single exponential decay of the population dynamics. 

\begin{figure*}[t]
	\begin{center}
		\includegraphics[trim=0mm 0mm 0mm 0mm, clip, width=2.15\columnwidth]{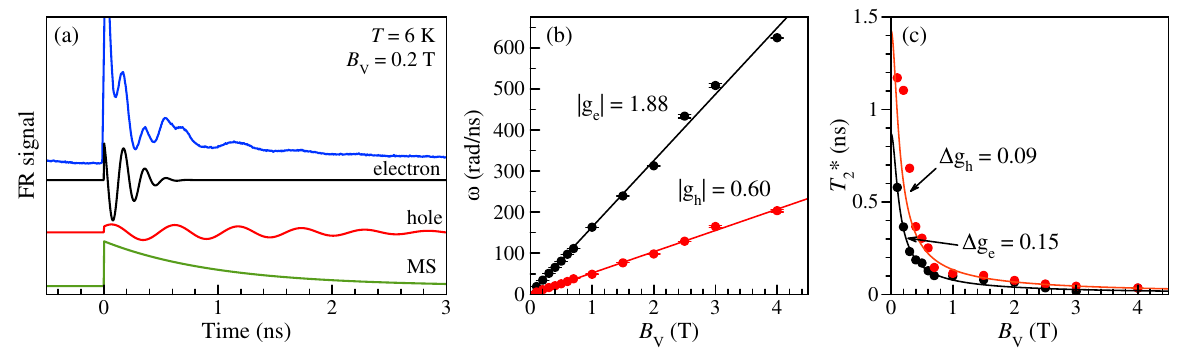}
		\caption{(a) Faraday rotation signal for $B_\text{V} = 0.2$\,T, $T = 6$\,K, and $E_\text{p} = 0.785$\,eV shown by blue trace. The pump and probe powers are 15\,mW and 1\,mW, respectively. The black and red traces are the electron and the hole spin contributions, respectively. The green trace is the signal of the molecular states. (b) The Larmor precession frequencies of electron (black) and hole (red) versus the transverse magnetic field ($B_\text{V}$). Linear fits give $|g_\text{e}| = 1.88$ and $|g_\text{h}| = 0.60$. (c) Dephasing times ($T_2^*$) of electron (black) and hole (red) as functions of $B_\text{V}$. The data are fitted using Eq.~(\ref{T2}).}
		\label{KR}
	\end{center}
\end{figure*}

An exemplary Faraday rotation signal from the QD ensemble at $B_\text{V} = 0.2$\,T is shown in Fig.~\ref{KR}(a) by the blue trace. We decompose it into three components, as shown by the colored traces below, using the following fit function:
\begin{equation}
\label{S}
S = \sum_i A_i\cos(\omega_it)\exp(-t^2/2T_{2,i}^{*2}).
\end{equation}
Here, the index $i$ gives the component number, $\omega_i = \mu_\text{B}g_i B_\text{V}/\hbar$ is the corresponding Larmor precession frequency, $g_i$ is the associated $g$-factor. $\mu_\text{B}$ is the Bohr magneton, and $\hbar$ is the reduced Planck constant. $T_{2,i}^{*}$ is the spin dephasing time.

The dependencies of the Larmor frequencies of electrons and holes on the transverse magnetic field ($B_\text{V}$) are linear, with the slopes being proportional to their $g$-factors. We obtain for the electron $|g_\text{e}| = 1.88$ and for the hole $|g_\text{h}| = 0.60$. These values agree with the data from Refs.~\cite{Belykh2015,Belykh2016}, where the assignment of the two contributions to electron and hole is done based on the differences in their $g$-factor anisotropies in magnetic field and corresponding theoretical calculations.

The spin dephasing times ($T_2^*$) of electrons and holes strongly depend on $B_\text{V}$ as shown in Fig.~\ref{KR}(c). At magnetic fields above 0.1\,T the spin dephasing is governed by the spread of the $g$-factor values, which is related to the inhomogeneities in the ensemble of QDs. At low magnetic fields, the spin dephasing saturates at the values determined by the hyperfine interaction with the fluctuating nuclear spin environment. As shown in Fig.~\ref{KR}(c), the hole spin dephasing saturates at a higher value of $T_2^*=1.4$\,ns than the electron with $T_2^*=0.6$\,ns. This can be explained by the weaker hyperfine coupling of the hole~\cite{TestelinPRB09,GlasenappPRB16}.

The dependence of $T_2^*$ on $B_\text{V}$ can, therefore, be described by:
\begin{equation}
\label{T2}
T_{2,i}^* = \hbar/\sqrt{(\Delta g_i \mu_\text{B}B_\text{V})^2 + (g_i \mu_\text{B}\Delta B_i)^2},
\end{equation}
where the index $i$ gives the electron or hole contribution, $\Delta g_i$ is the spread of $g$ factors, $\Delta B_i$ is the limiting factor at low fields~\cite{Mikhailov2018}. The fitting curves for electron and hole are shown in Fig.~\ref{KR}(c) by the solid lines. The evaluation gives $\Delta g_\text{e} = 0.15$, $\Delta g_\text{h} = 0.09$, and $\Delta B_\text{e} = 7$\,mT, $\Delta B_\text{h} = 13$\,mT. The reported $g$-factor dispersions exceed the previously reported $\Delta g_\text{e} = 0.002$, $\Delta g_\text{h} = 0.05$ for  InAs/InAlGaAs QDs~\cite{Belykh2015}, due to the more complex design of the QD structure involving the thin barrier between the QD layers.

\begin{figure}[b]
	\begin{center}
		\includegraphics[trim=0mm 0mm 0mm 0mm, clip, width = 1.0\columnwidth]{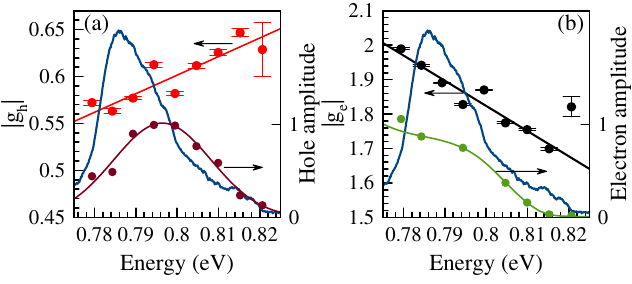}
		\caption{(a) Red dots present the dependence of $|g_\text{h}|$ on the central laser energy $E_\text{p}$. The red line is a linear fit to the data. The brown dots show the spectral dependence of the normalized hole spin amplitude contribution to the FR signal. The solid line is a guide for the eye. (b) Black dots present the dependence of $|g_\text{e}|$ on $E_\text{p}$. The black solid line is a linear fit. The green dots show the spectral dependence of the normalized electron contribution. The solid line is a guide for the eye.}
		\label{SD}
	\end{center}
\end{figure}

The spread of the parameters in the ensemble of QDs manifests itself in the broadening of the PL line. Despite that, the electron and hole $g$-factors show a systematic variation throughout the range of optical transition energies in the ensemble, as was earlier observed for near-infrared QDs and telecom range QDs~\cite{Belykh2015}. By using the spectrally narrow laser, one can select and test QD subensembles by tuning the central energy $E_\text{p}$. The dependence of $|g_\text{h}|$ on the pump-probe energy is presented in Fig.~\ref{SD}(a), with the absolute value increasing with $E_\text{p}$, as shown by the red dots. The dependence can be described by the linear function $|g_\text{h}| = AE_\text{p} - B$, with $A = 2\times 10^{-3}$\,meV$^{-1}$ and $B = 1\times 10^{-3}$. The electron g factor $|g_\text{e}|$ shows the opposite tendency: the absolute value is decreasing with $E_\text{p}$, as shown in Fig.~\ref{SD}(b). It can be described by $|g_\text{e}| = -A E_\text{p} + B$, with $A = 7.3\times 10^{-3}$\,meV$^{-1}$ and $B = 7.7\times 10^{-3}$. According to the Roth-Lax-Zwerdling equation, the increase of the absolute value of the electron $g$-factor shows a negative sign~\cite{Belykh2015}. The hole should have a positive $g$-factor, opposite to the electron. If we now compare the amplitude of the signal contributions, one can see that the amplitude of the Faraday rotation signal for the hole has a pronounced resonance with the maximum at $E_\text{p} = 0.796$\,eV (see Fig.~\ref{SD}(a)). By contrast, the electron amplitude constantly increases toward the low-energy side of PL, as shown in Fig.~\ref{SD}(b). We concentrate on the spectral position with the highest hole spin amplitude for further measurements. 

Finally, Fig.~\ref{KR}(a) demonstrates an additional, non-oscillating decaying signal component (MS), despite the applied external magnetic field. It arises from the complexes with zero total spin configuration in the singlet and triplet ($T_0$) molecular states of paired carriers (preferentially electrons), residing in neighboring dots. This property requires further investigations, including studies of the influence of the barrier thickness and the type of resident charging. This is a topic for future studies, and here, we are focusing on the signal from the carriers localized in a QD, similar to isolated QDs.

\subsection{Longitudinal spin relaxation}

\begin{figure*}[t]
	\begin{center}
		\includegraphics[trim=0mm 0mm 0mm 0mm, clip, width=2.0\columnwidth]{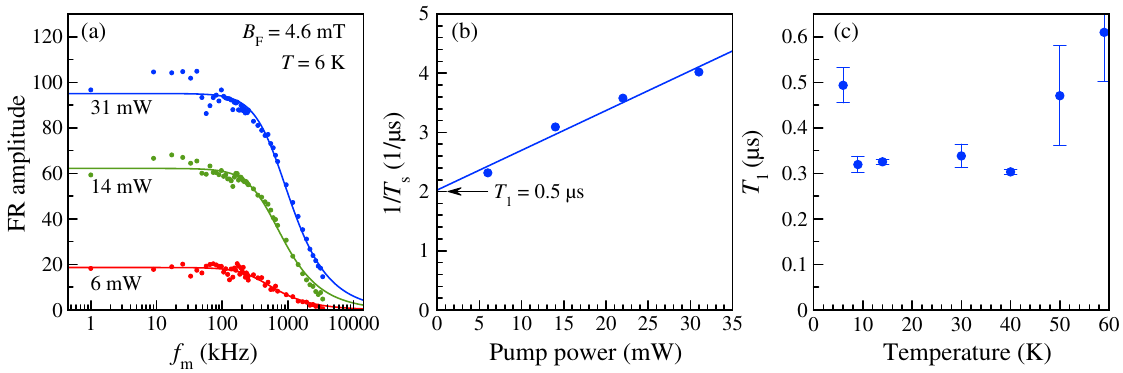}
		\caption{(a) Faraday rotation amplitude (dots) at the pump-probe delay of $-50$\,ps as a function of the pump modulation frequency $f_\text{m}$ measured in a longitudinal magnetic field of $B_\text{F} = 4.6$\,mT for different pump powers. $E_\text{p} = 0.784$~eV. The lines are fita according to Eq.~(\ref{eq_SIN}) describing the spin inertia effect. (b) Power dependence of inverse spin lifetimes $1/T_\text{s}$. A linear extrapolation to zero power gives the intrinsic spin relaxation time $T_1 = 0.5$\,$\mu$s. (c) $T_1$ dependence on temperature.}
		\label{SIN}
	\end{center}
\end{figure*}

To measure the longitudinal spin lifetime ($T_\text{s}$) and the spin relaxation time ($T_1$) of the holes, we employ the spin inertia technique~\cite{Heisterkamp2015}. The helicity of the pump is modulated between $\sigma^+$ and $\sigma^-$ with the frequency $f_\text{m}$. When the modulation period becomes shorter than the spin lifetime $T_\text{s}$, the averaged signal amplitude starts to decrease. The measurement is done at negative time delays and in a weak longitudinal magnetic field ($B_\text{F}$), to avoid the influence of the nuclear spin fluctuations.  

The spin inertia curves are shown in Fig.~\ref{SIN}(a) at $B_\text{F} = 4.6$\,mT for different pump powers. The decay of the FR signal in dependence on $f_\text{m}$ can be described by:
\begin{equation}
\label{eq_SIN}
S(f_\text{m}) = S_0/\sqrt{1 + (2\pi f_\text{m}T_\text{s})^2},
\end{equation}
with $T_\text{s}  = 0.3$\,$\mu$s at the pump power of 14\,mW. The extrapolation of $T_\text{s}$ dependence on power to zero gives the spin relaxation time $T_1 = 0.5$\,$\mu$s, see Fig.~\ref{SIN}(b). The experimental data can be well-fitted by this form, indicating spin relaxation times exceeding the laser repetition period. This allows us to attribute this signal to resident carrier spins. 

The temperature stability of the spin lifetime up to 50\,K, shown in Fig.~\ref{SIN}(c), confirms the observations reported earlier for similar types of QDs~\cite{Mikhailov2018}.

\subsection{Transverse spin relaxation}

The large spread of $g$-factors and long spin relaxation times are promising conditions for the spin mode-locking effect (SML), observed on electrons in n-doped (In,Ga)As/GaAs QDs~\cite{Greilich2006Sci,Evers2018}, on holes in p-doped InAs/GaAs QDs~\cite{Fras2012,Varwig2012}, and recently on holes in CsPb(Cl,Br)$_3$ perovskite nanocrystals~\cite{Kirstein_ML}. In this paper, we show the spin mode-locking effect for the holes in the QDs ensemble emitting in the telecom range. 

\begin{figure*}[t]
	\begin{center}
		\includegraphics[trim=0mm 0mm 0mm 0mm, clip, width = 2.1\columnwidth]{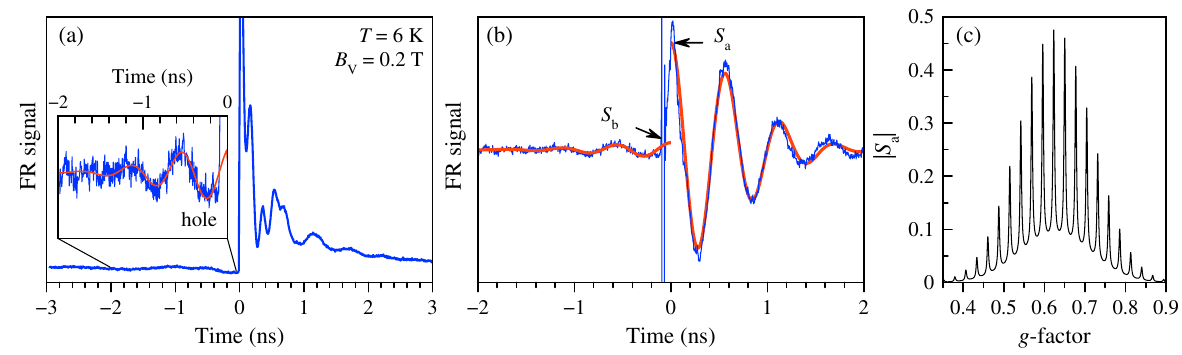}
		\caption{(a) Faraday rotation signal measured at $B_\text{V} = 0.2$\,T, $T = 6$\,K and $E_\text{p} = 0.785$\,eV. The pump and probe powers are 15\,mW and 1\,mW. The inset shows the zoomed-in signal at negative time delays. (b) The blue trace is the spin mode-locking signal of the holes, and the red solid lines shows its modeling. (c) Spectral distribution of the precession modes calculated for $T_2 = 400$\,ns and $\Theta=0.6\pi$.}
		\label{ML}
	\end{center}
\end{figure*}

The resident hole spin polarization is created through the trion excitation mechanism in the pump-probe Faraday rotation~\cite{Yugova2009}. The polarized spins precess about $B_\text{V}$ and loose their polarization with the spin coherence time of a single carrier $T_2$. After the action of an infinite number of pulses with a repetition period of $T_\text{R}$, the spin polarization accumulates for carriers with $T_2 > T_\text{R}$. In the inhomogeneous ensemble of QDs, the sum of the multiple oscillating signals with Larmor frequencies commensurate with $\omega_\text{R}$ contributes to the SML signal. It manifests itself in the revival of the FR signal at negative delays (probe arrives before pump). Such a signal is shown in the inset of Fig.~\ref{ML}(a). To assign the carrier contribution to it, we measure the SML dependence on $B_\text{V}$. From the $g$-factor values, we conclude that the signal is related to the hole spins. 

To evaluate the spin coherence time $T_2$, we have decomposed the signal into its components. Figure~\ref{ML}(b) demonstrates the remaining signal component after subtraction of the non-oscillating decaying component, the electron signal, and the hole signal with short dephasing times at positive time delays. The signal decay time at negative and positive delay times is given by the spin dephasing time $T_2^*$, which is limited by the spread of $g$-factors, as discussed in section~III.A. The relation of the signal amplitude before the pump pulse arrival $S_\text{b}$ to the amplitude after the pulse action $S_\text{a}$ is determined by $T_2$ and the pump power, and allows one to extract the corresponding time. 

To estimate the $T_2$, we use the model developed in Refs.~\cite{Yugova2009,Yugova2012}. Each optical pulse changes the spin polarization proportional to pump pulse area $\Theta$ (or the pump power in the experiment):
\begin{eqnarray}
\label{OPz}
S_\text{z,a} = -\frac{\sin^2(\Theta/2)}{4} + \frac{\cos^2(\Theta/2)  + 1}{2}S_\text{z,b},\\
S_\text{y,a} = \cos(\Theta/2)S_\text{y,b},\\
\label{OPx}
S_\text{x,a} = \cos(\Theta/2)S_\text{x,b}.
\end{eqnarray}
Here, the $z$-component is directed along the $\textbf{k}$-wavevector of light, and the $x$-component is along $B_\text{V}$. Equations~\eqref{OPz}-\eqref{OPx} assume rectangular optical pulses and do not include the optical detuning of the laser energy from the trion resonance.

The dynamics of the hole spin polarization after optical excitation is determined by the Bloch equations:
\begin{eqnarray}
S_\text{z}(t) = [S_\text{z,a}\cos(\omega t) + S_\text{y,a}\sin(\omega t)]\exp(-t/T_2),\\
S_\text{y}(t) = [S_\text{y,a}\cos(\omega t)-S_\text{z,a}\sin(\omega t)]\exp(-t/T_2),\\
S_\text{x}(t) = S_\text{x,a}\exp(-t/T_1).
\end{eqnarray}
Note that the $S_\text{x}$ component cannot be created without optically detuned optical pulses. We consider the spin relaxation time anisotropy in the sample plane.

The accumulation of the spin polarization for an infinite number of optical pulse actions is calculated numerically for each individual spin in the ensemble. We use a Gaussian distribution of the $g$-factors in the ensemble with the width of $2\Delta g_\text{h} = 0.2$. The resulting precession modes are shown in Fig.~\ref{ML}(c). The width of each precession mode as well as $S_\text{b}/S_\text{a}$ are determined by $T_2$ and $\Theta$. For the calculations, we use the experimentally measured $g_\text{h}$, $\Delta g_\text{h}$, the result is shown in Fig.~\ref{ML}(b) by the red line. 

The parameter $\Theta$ can not be precisely evaluated from the experiment because the SML amplitude neither shows amplitude saturation nor Rabi oscillations in dependence on the pump power due to the high inhomogeneity of the QD ensemble. We, therefore, have to consider the range of pulse areas with $\Theta=0.6\pi$, leading to the maximal $T_2 = 400$\,ns, and the $\Theta = \pi$ with the minimal possible $T_2=20$\,ns, both of which lead to the observed amplitude relation $S_\text{b}/S_\text{a}$.

Being initially n-doped by the Si $\delta$-doped layer, the QDs are expected to show a long spin relaxation time for the resident electrons. However, we observe that the hole spin demonstrates extended spin coherence and relaxation, as determined by the spin mode-locking and spin-inertia methods, respectively.

We conclude that the hole spin is created through optical excitation in a QD and resides in the same QD without tunneling. As was previously studied for InAs QD molecules, the tunnel barriers with width on the order of 15\,nm lead to a strongly reduced tunneling probability for holes due to their increased effective mass in comparison to electrons~\cite{BrackerAPL07}.

The detected spin coherence is comparable with the one observed in InAs QDs emitting in the range of 0.9\,$\mu$m, demonstrating $T_2\approx 1\,\mu$s~\cite{VarwigPRB14,HutmacherPRB18}. However, to be useful for QR technology, we must reach spin coherence times close to milliseconds (100\,km is about 0.33\,ms for light traveling). Such times could be achieved by using the coherence transfer to the nuclear surrounding in the QDs, showing coherence times in the required time range~\cite{WüstNatNanotech16,Millington-HotzeNatComm23}. The feasibility of such protocol is currently investigated~\cite{BodeyNPJQI19,GangloffScience19}.

\section{Conclusions}

Using the spin mode-locking method, we could determine the range of the achievable spin coherence of holes, which exceed previously measured values in similar types of QDs by at least an order of magnitude~\cite{Mikhailov2018}. The longitudinal spin lifetime of 0.5\,$\mu$s is also extended and currently provides the upper limit for the spin coherence. Observing a non-oscillating signal in transverse magnetic fields confirms the formation of molecular states and requires further investigation.

\section{Acknowledgements}
We acknowledge the financial support by the Deutsche Forschungsgemeinschaft in the frame of the ICRC TRR 160 (Project A1) and DFG-Heisenberg grant-BE 5778/4-1. We also acknowledge the Bundesministerium f\"{u}r Bildung und Forschung in the frame of the project QR.X (Contract No. 16KISQ011 and 16KISQ005).

%

\end{document}